\documentclass[sigconf]{acmart}
\AtBeginDocument{%
  }

\setcopyright{acmlicensed}
\copyrightyear{2026}
\acmYear{2026}
\setcopyright{cc}
\setcctype{by}
\acmConference[KDD 2026] {Proceedings of the 32nd ACM SIGKDD Conference on Knowledge Discovery and Data Mining V.1}{August 9--13, 2026}{Jeju Island, Republic of Korea.}
\acmBooktitle{Proceedings of the 32nd ACM SIGKDD Conference on Knowledge Discovery and Data Mining V.1 (KDD 2026), August 9--13, 2026, Jeju Island, Republic of Korea}
\acmISBN{979-8-4007-2258-5/2026/08}
\acmDOI{10.1145/3770854.3780272}

\usepackage{algorithm}
\usepackage{algpseudocode}
\usepackage{amsmath}
\usepackage{xspace}
\usepackage{booktabs}
\usepackage{color,soul}
\usepackage{wasysym}
\sethlcolor{yellow}
\usepackage{multirow}
\usepackage{enumitem}

\begin{document}

\title{SEW: Strengthening Robustness of Black-box DNN Watermarking via Specificity Enhancement}


\author{Huming Qiu}
\affiliation{%
  \institution{Fudan University}
  \state{Shanghai}
  \country{China}}
\email{hmqiu23@m.fudan.edu.cn}

\author{Mi Zhang}
\authornote{Corresponding authors.}
\affiliation{%
  \institution{Fudan University}
  \state{Shanghai}
  \country{China}}
\email{mi_zhang@.fudan.edu.cn}

\author{Junjie Sun}
\affiliation{%
  \institution{Alibaba Group}
  \state{Beijing}
  \country{China}}
\email{sunjunjie.sjj@alibaba-inc.com}

\author{Peiyi Chen}
\affiliation{%
  \institution{Fudan University}
  \state{Shanghai}
  \country{China}}
\email{peiyichen21@m.fudan.edu.cn}

\author{Xiaohan Zhang}
\authornotemark[1]
\affiliation{%
  \institution{Fudan University}
  \state{Shanghai}
  \country{China}}
\email{xh_zhang@fudan.edu.cn}

\author{Min Yang}
\affiliation{%
  \institution{Fudan University}
  \state{Shanghai}
  \country{China}}
\email{m_yang@fudan.edu.cn}

\renewcommand{\shortauthors}{Huming Qiu et al.}

\newcommand{\metricname}{specificity\xspace}
\newcommand{\wmname}{\textsc{SEW}\xspace}
\newcommand{\wmallname}{Specificity-Enhanced Watermarking\xspace}


\begin{abstract}
To ensure the responsible distribution and use of open-source deep neural networks (DNNs), DNN watermarking has become a crucial technique to trace and verify unauthorized model replication or misuse. In practice, black-box watermarks manifest as specific predictive behaviors for specially crafted samples. However, due to the generalization nature of DNNs, the keys to extracting the watermark message are not unique, which would provide attackers with more opportunities. Advanced attack techniques can reverse-engineer approximate replacements for the original watermark keys, enabling subsequent watermark removal.
In this paper, we explore black-box DNN watermarking \textit{specificity}, which refers to the accuracy of a watermark's response to a key. Using this concept, we introduce \textit{Specificity-Enhanced Watermarking} (SEW), a new method that improves specificity by reducing the association between the watermark and approximate keys. Through extensive evaluation using three popular watermarking benchmarks, we validate that enhancing specificity significantly contributes to strengthening robustness against removal attacks. SEW effectively defends against six state-of-the-art removal attacks, while maintaining model usability and watermark verification performance.
\end{abstract}




\maketitle

\section{Introduction}
Deep Neural Networks (DNNs) has revolutionized various fields including computer vision \cite{parkhi2015deep, he2016deep}, natural language processing \cite{vaswani2017attention, devlin2018bert}, and recommendation systems \cite{zhang2019deep}.
In recent times, HuggingFace \cite{huggingface2023openrail, stable-diffusion-license} and other model vendors \cite{BigScienceRAILM2022} have started releasing open-source models under the OpenRAIL-M series license. This license requires that open-source models and their derivatives be accompanied by the license, with the goal of promoting responsible distribution and usage of these models \cite{RAIL2022}.
To prevent unauthorized replication or malicious misuse of open-source models, DNN watermarking has emerged as a new tool for tracing and verifying model intellectual property \cite{boenisch2021systematic}.

\begin{figure}
\centering
\includegraphics[trim=0 0 0 0,clip,width=0.45\textwidth]{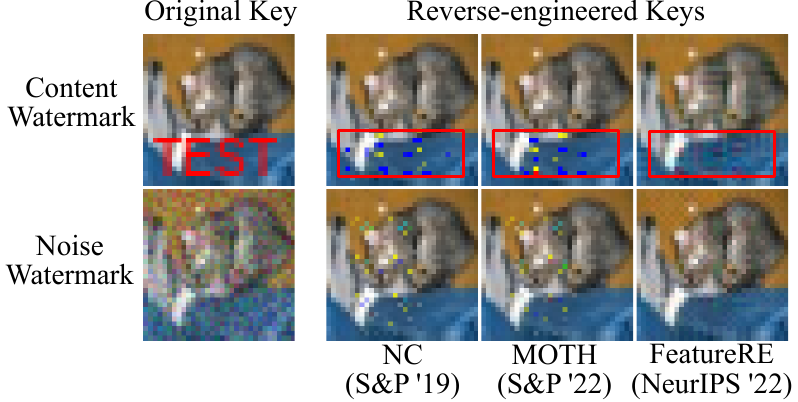}
\caption{Examples of reverse-engineered keys generated by watermark removal attacks. The first column shows the original key with "TEST" text and full-image noise, while the last three columns display the reverse-engineered keys.
}
\label{fig:reverse_trigger}
\end{figure}

DNN watermarking embeds copyright information (i.e., watermarks) by manipulating the model internals (i.e., \textit{white-box}) or the prediction behaviors (i.e., \textit{black-box}) \cite{li2021survey}. The watermarks can later be extracted using specific watermark keys, creating an invisible mechanism for protecting intellectual property.
Black-box watermarking \cite{gao2020backdoor} is a technique for embedding additional functionality in DNNs by modifying the training data. It aims to map inputs containing a specific trigger (i.e., \textit{watermark key}) to a designated label (i.e., \textit{watermark message}).
Black-box watermarking only requires black-box access to the model, which is more practical for watermark verification \cite{lu2023mira} and is the main focus of this paper.

However, current black-box watermarking schemes lack sufficient robustness to withstand various attack patterns \cite{lukas2022sok, lee2022evaluating}.
Black-box watermarking can be viewed as a benign application of backdoor attacks \cite{adi2018turning}, and attackers have exploited backdoor countermeasures to remove watermarks \cite{aiken2021neural, wang2019neural}.
The watermark keys are inaccessible to attackers, but many removal attacks attempt to find a substitute for the watermark key to facilitate watermark removal \cite{tao2022better, wang2022rethinking}. For example, a recent attack can remove the watermark by relearning reverse keys \cite{lu2024neural}. As shown in \autoref{fig:reverse_trigger}, while these identified keys for removal attacks may not be the exact original keys, they still have the ability to extract watermark message and can be considered as approximate keys.

\noindent{\bf Our Work.} Black-box watermarking is like a lock, where the watermark key serves as the key to unlock it, and the watermark verification process is similar to pairing a lock and its key. Because of the generalization property of DNNs, there are multiple watermark keys that can unlock the watermark. This gives attackers more opportunities to exploit, which is why removal attacks are effective. We term the precision of watermark-key responses as watermark \metricname and designate keys that differ from the original key yet are still capable of extracting watermark message as approximate keys.
Given the prevalence of approximate keys and their significant success in watermark removal attacks, a natural question arises: \textit{can we suppress the existence of approximate keys to withstand state-of-the-art watermark removal attacks?}

In this paper, we provide an affirmative answer by revealing the positive effect of watermark specificity on strengthening robustness.
We first delve into the intimate interplay between watermark \metricname and approximate keys, devising a effective metric for measuring watermark \metricname. Specifically, we conduct a comprehensive noise analysis on watermark keys, estimating the noise boundary that maintains the effectiveness of the keys. This boundary delineates the potential range of existence for approximate keys.
Leveraging this boundary, we introduce, for the first time, a metric to gauge watermark \metricname. Furthermore, we assess the \metricname of existing black-box watermarks based on this metric, revealing subpar performance in \metricname for traditional black-box watermarking, thereby compromising their robustness against removal attacks. These findings offer potent insights for research and improvement in black-box DNN watermarking techniques.

To this end, guided by watermark \metricname, we propose a novel black-box watermarking technique named \wmallname (\wmname). \wmname is designed to refine watermark keys, precisely defining the conditions for watermark response, thereby rendering approximate keys incapable of extracting watermark message and enhancing the \metricname of the watermark. Specifically, during the watermark embedding phase, we create two sets of trigger samples carrying keys. One set of samples carries the original key, with labels modified to the target label, aiming to embed the watermark into the model. The other set of samples carries approximate keys, while still retaining the true labels, aiming to suppress the association between the watermark and approximate keys, thus reinforcing the watermark \metricname.

In summary, we mainly make the following key contributions:

\begin{itemize}[leftmargin=10pt, noitemsep, nosep]
\item We delve into the specificity of black-box DNN watermarks, which reveals \textit{for the first time} the positive effect of specificity on robustness. By devising the algorithm to quantify specificity, we provide a novel and practical perspective for understanding the robustness of existing black-box watermarks.

\item Building on the watermark \metricname, we propose a new black-box DNN watermarking scheme, \wmallname (\wmname), designed to reduce the correlation between watermarks and approximate keys, thereby strengthening robustness against removal attacks.

\item We conduct a comprehensive evaluation of the performance of \wmname across six state-of-the-art removal attacks, demonstrating its efficacy in defending against these attacks while preserving model performance and watermark verification rates. This further supports our observation that highly specific watermarks exhibit stronger robustness. To facilitate future studies, we open-source our code in the following repository: \url{https://huggingface.co/Violette-py/SEW}.
\end{itemize}

\section{Related Work}

\subsection{Black-box DNN watermarking}

As a method for tracing ownership of DNN models, DNN watermarking technology aims to embed watermarking functionality into the model, with only the watermark key held by the model owner capable of extracting the hidden watermark \cite{sun4697020deep}. Depending on the level of access required for the watermark verification process, DNN watermarking can be categorized into white-box watermarks and black-box watermarking. White-box watermarks typically embed watermark message into internal aspects of the model, such as model parameters \cite{uchida2017embedding, wang2021riga}, specific structures \cite{chen2021you}, or neuron activations \cite{darvish2019deepsigns}. In contrast, the embedding process of black-box watermarking resembles a backdoor injection process \cite{hua2023deep}, whereby specific input-output pairs are constructed during training to compel the model to learn a secret additional functionality, thus enabling ownership verification without accessing the model's internal information \cite{chen2019blackmarks}.

Early black-box watermarking methods mainly achieve watermark embedding by introducing triggers or use out-of-distribution samples \cite{zhang2018protecting}. Subsequent studies further extend trigger design and embedding mechanisms to improve watermark robustness, including entangled watermark embedding based on the soft nearest neighbor loss \cite{jia2021entangled}, approaches that construct key datasets using adversarial examples or clean images \cite{le2020adversarial,namba2019robust}, and signature-based pixel-level watermark modification strategies \cite{guo2018watermarking}. Recent work explicitly enlarges the decision margins of watermarked samples, which effectively enhances the model’s robustness against model stealing \cite{kim2023margin}. In addition, some emerging studies move beyond traditional prediction-based watermarking paradigms and embed richer watermark information into model behavior, encoding multi-bit watermarks through feature attribution explanations \cite{shao2024explanation}.

To ensure reliable ownership verification, an ideal black-box DNN watermarking scheme should meet the following requirements \cite{yan2023rethinking}:
\noindent{\bf (\romannumeral 1) Fidelity.} The watermark should not or minimally affect the model's usability.
\noindent{\bf (\romannumeral 2) Robustness.} The watermark should withstand potential removal attacks.
\noindent{\bf (\romannumeral 3) Integrity.} Independently trained models should not be identified as containing the watermark.
\noindent{\bf (\romannumeral 4) Specificity.} The watermark message should only be extracted by the original key.
Notably, existing black-box watermarking often ignore the specificity when designing, and its impact on robustness has not been thoroughly investigated \cite{chen2024deep}. To address this research gap, this paper focuses on watermark specificity.

\subsection{Watermark Removal Attacks}

Black-box DNN watermarking establishes a connection between specified data and target labels, essentially constituting a benign application of backdoor attacks, naturally inheriting the associated vulnerabilities of backdoors \cite{shafieinejad2021robustness}. Consequently, some backdoor defense methods have been exploited by attackers for watermark removal attacks \cite{wang2019neural}. Generally, existing watermark removal attacks can be broadly categorized into two types \cite{lukas2022sok}: model extraction attacks and model modification attacks.

Model extraction attacks aim to observe the input-output behavior of a model to infer its internal structure and parameter information \cite{takemura2020model}. Typically, extraction attacks are employed in black-box scenarios where attackers cannot directly access the target model's parameters or structural information \cite{orekondy2019knockoff}. Instead, they need to gradually infer the internal workings of the model through a large number of queries and feedback \cite{shafieinejad2021robustness}, thereby achieving model extraction.
For instance, in the case of hard-label extraction attacks \cite{tramer2016stealing}, attackers utilize the watermarked model to assign predicted labels to query data and train a proxy model without the watermark on this pseudo-labeled dataset.

Model modification attacks require white-box access to the target model, allowing attackers to remove the watermark with minimal cost. Modified attacks are categorized into three types \cite{sun2021detect}: (\romannumeral 1) Pruning-based Attacks. Redundant weights or neurons \cite{liu2018fine} in the pruned model are removed to render the underlying watermark ineffective. (\romannumeral 2) Finetuning-based Attacks. Carefully designed fine-tuning techniques \cite{chen2019leveraging} are employed to train the model for several additional epochs to eliminate the watermark. (\romannumeral 3) Unlearning-based Attacks. These aim to obtain approximate keys through reverse engineering methods \cite{lu2024neural}, thereby prompting the model to forget the association between the watermark and the key.

The attack budgets required for the above attacks are summarized in \autoref{tab:attack budgets}. For example, extraction attacks impose the lowest requirements on access privileges, but they typically rely on extensive access to in-distribution data and incur high training costs, since the model often needs to be retrained from scratch to ensure usability. Although some studies show that the data access requirements of model stealing can be reduced by using out-of-distribution data or AI-generated synthetic data \cite{orekondy2019knockoff}, the overall data demand remains significantly higher than that of other attacks. In contrast, modification attacks require white-box access to the model and therefore demand stricter access privileges. In this setting, an adversary can remove the embedded watermark through fine-tuning or pruning with a small amount of clean data, which substantially reduces training cost while preserving model usability.

\begin{table}[t]
\centering
\caption{Comparison of attack budgets among existing watermark removal attacks, where $\Circle/\LEFTcircle/\CIRCLE$ denotes low (or none), medium, and high attack budgets.}
\resizebox{0.4 \textwidth}{!}
{
    \begin{tabular}{c c c c c}
    \toprule
    \textbf{Type} & \textbf{\begin{tabular}[c]{@{}c@{}}Access \\ Permissions\end{tabular}} & \textbf{\begin{tabular}[c]{@{}c@{}}Dataset \\ Access\end{tabular}} & \textbf{\begin{tabular}[c]{@{}c@{}}Training \\ Cost\end{tabular}}& \textbf{\begin{tabular}[c]{@{}c@{}}Utility \\ Loss\end{tabular}} \\
    \midrule
    \textbf{\begin{tabular}[c]{@{}c@{}}Extraction\end{tabular}} & $\Circle$ & $\CIRCLE$ & $\CIRCLE$ & $\Circle$\\
    \textbf{\begin{tabular}[c]{@{}c@{}}Prune\end{tabular}} & $\CIRCLE$ & $\Circle$ & $\Circle$ & $\LEFTcircle$\\
    \textbf{\begin{tabular}[c]{@{}c@{}}Finetune\end{tabular}} & $\CIRCLE$ & $\LEFTcircle$ & $\LEFTcircle$ & $\LEFTcircle$\\
    \textbf{\begin{tabular}[c]{@{}c@{}}Unlearn\end{tabular}} & $\CIRCLE$ & $\Circle$ & $\LEFTcircle$ & $\Circle$\\
    \bottomrule
    \end{tabular}
}
\label{tab:attack budgets}
\end{table}

\section{SECURITY SETTINGS}
Recent studies have systematically evaluated the robustness of black-box watermarking when facing watermark removal attacks, concluding that its robustness has yet to meet the requirements, and there does not exist a robust watermarking scheme capable of completely thwarting all attacks \cite{lukas2022sok, lu2023mira}.
Given that our work aims to enhance the robustness of open-source model watermarks, in \autoref{sec:threat_model}, we hypothesize a threat model wherein adversaries attempt to remove the watermark embedded within open-source models. Additionally, in \autoref{sec:limitations}, we discuss the limitations of existing black-box watermarking against extraction attacks.

\subsection{Threat Model} \label{sec:threat_model}
\noindent{\bf Attack Scenarios and Capabilities.} In our threat model, adversary $\mathcal{A}$ acquires an open-source model $f_w$ from popular platforms like HuggingFace and attempts to violate the terms of the OpenRAIL-M license by maliciously exploiting $f_w$, either in open-source or closed-source fashion. Similar to settings in recent watermark robustness studies \cite{lee2022evaluating, lukas2022sok}, $\mathcal{A}$ possesses white-box privileges to access the internal parameters of $f_w$ and has the capability to modify them. $\mathcal{A}$ is aware of the existence of a watermark within $f_w$ used for tracing and verifying model IP. Therefore, $\mathcal{A}$'s objective is to derive a surrogate model $f_a$ from $f_w$ without the watermark, thereby circumventing ownership verification.

\noindent{\bf Defense Goals.} The aim of this paper is to enhance the robustness of existing black-box watermarks, thereby facilitating model supply platforms to robustly trace open-source models in a black-box manner, thus preventing malicious misuse or irresponsible distribution.
A robust watermark should possess the ability to resist watermark removal attacks and survive within the surrogate model $f_a$. Even if adversaries manage to remove the watermark through more potent attacks, it inevitably comes with a loss of model usability \cite{yang2021robust}.
Adversary $\mathcal{A}$ may maliciously exploit the surrogate model $f_a$ in a closed-source manner. Black-box watermarking enables IP verification personnel to extract watermark message as a hallmark of copyright declaration by simply utilizing a trigger set carrying watermark keys as queries for $f_a$, without needing to access the internal information of $f_a$.

\subsection{Limitations of Extraction Attacks} \label{sec:limitations}

Extraction-based attacks rely solely on black-box access to the model API, as summarized in \autoref{tab:attack budgets}. While this may seem advantageous in terms of accessibility, it imposes substantial overhead on adversaries, who must train surrogate models from scratch and obtain a large volume of in-distribution query data \cite{shafieinejad2021robustness}. In practice, adversaries with access to such extensive resources are more likely to develop their own models independently rather than risk legal or technical complications by misappropriating a pre-trained model.
In contrast, model modification attacks operate under white-box assumptions and require significantly lower costs. With direct access to model parameters and minimal data collection, adversaries can efficiently remove watermarks using techniques such as fine-tuning, pruning, or targeted unlearning \cite{wang2019neural, wang2022rethinking, tao2022model}. Notably, our threat model focuses on open-source model usage and redistribution, where white-box access is inherently available. Under these realistic conditions, modification-based attacks present a more practical and economically feasible threat vector. As such, we exclude extraction attacks from our evaluation scope due to their limited applicability and high cost under our assumed threat model.

Instead, we focus on model modification attacks, particularly Dehydra \cite{lu2024neural}, a recent and specialized approach tailored for watermark removal. Unlike general-purpose backdoor defenses, Dehydra is explicitly designed to exploit the association between watermark keys and model outputs, making it highly effective in removing embedded watermarks. This focus underscores the conceptual and technical differences between watermarks and conventional backdoors, and ensures that our evaluation accurately reflects the specific challenges of watermark robustness in real-world scenarios.

\section{Methodology}

\subsection{Watermark Specificity Measurement}
For measuring watermark specificity, computing the noise boundary that maintains the effectiveness of keys is the algorithm's primary step. However, accurately computing this noise boundary faces numerous challenges \cite{katz2017reluplex}. Firstly, the complexity and non-linear characteristics of deep learning models increase the computational cost and complexity of calculating the noise boundary, making it difficult to directly establish mathematical models. Secondly, precise computation of the noise boundary involves complex mathematical forms and high-dimensional spaces, making it challenging to obtain accurate closed-form solutions through analytical methods. Additionally, different model architectures and watermarking schemes may require customized methods for computing the noise boundary, adding to the diversity of computations.

To address these challenges, we convert the computational problem for specificity into an optimization problem for noise upper bounds.
The core idea of the measurement algorithm is to compute the maximum noise bound that preserves the effectiveness of the key, which indicates the potential range of approximate keys. A smaller range means a lesser potential number of approximate keys, which indicates a higher specificity of the watermark.
By formulating an optimization function, we can search for the maximum intensity of noise while ensuring that the key can still activate the watermark after enduring this noise. The advantage of this optimization approach lies in its ability to iteratively approximate the noise boundary, thereby better adapting to challenges in different scenarios, including various neural network architectures and diverse watermark designs. This provides us with a universal method for effectively evaluating and measuring watermark specificity.

\begin{figure}
    \centering
    \includegraphics[width=0.75\columnwidth]{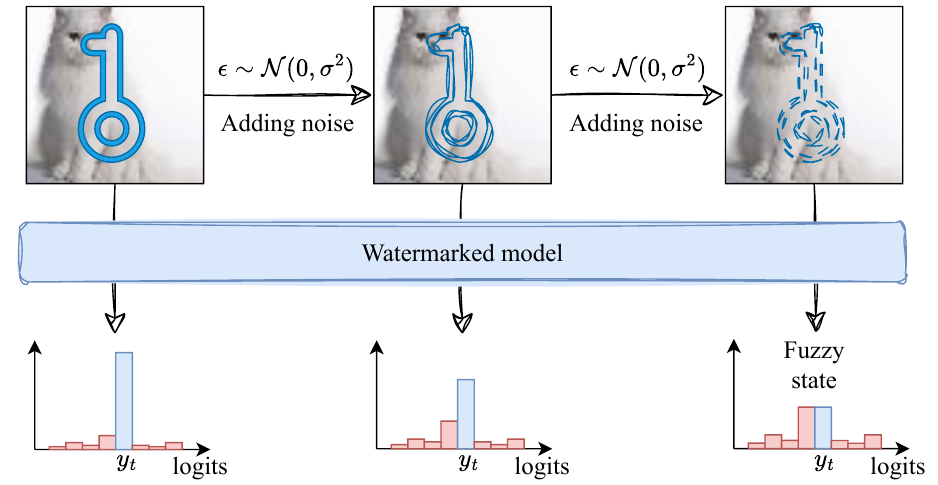}
    \caption{Noise addition analysis process for samples with the watermark key, the fuzzy state indicate being in the noise upper bound.
    }
    \label{fig:noise_ans}
\end{figure}

\begin{figure*}
    \centering
    \includegraphics[width=1.6\columnwidth]{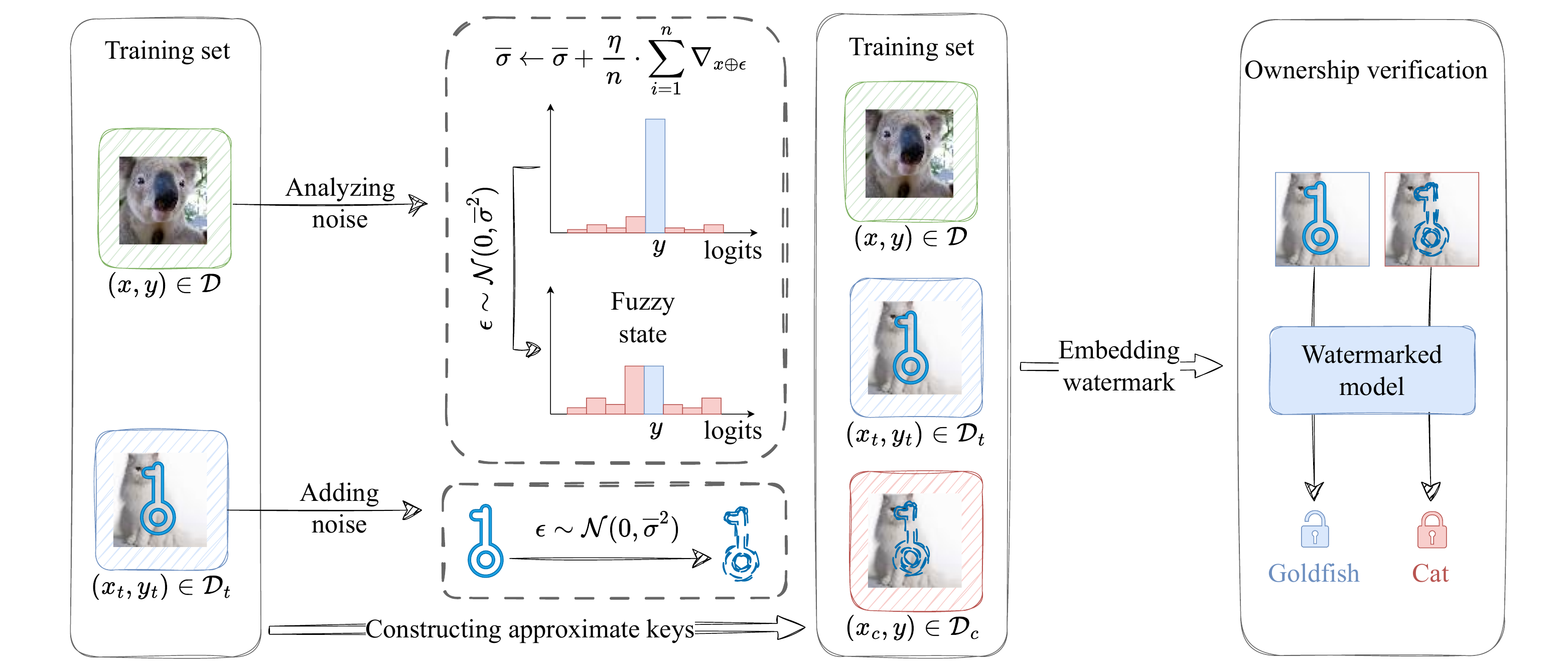}
        \caption{Overview of \wmname.}
    \label{fig:overview}
\end{figure*}

Formally, let us consider a classical image classification problem with $k$ classes, where the samples $x \in \mathbb{R}^d$ and the corresponding labels $y \in \{0, 1, \ldots, k\}$ follow the joint distribution $D(x, y)$. A neural network $F_\theta : \mathbb{R}^d \rightarrow \{0, 1, \ldots, k\}$ with parameters $\theta$ trained on this distribution should satisfy $\arg\max_\theta \mathbb{P}(x,y) \sim D[F_\theta(x) = y]$.

\begin{definition}\label{def:one}
(Noisy Key).
Given a Gaussian noise $\epsilon_\sigma \sim \mathcal{N}(0, \sigma^2)$ with $\sigma$ standard deviation and an original key $t$. If $\hat{t} = t \oplus \epsilon_\sigma$ and $\sigma \neq 0$, we say that $\hat{t}$ is a noisy key with noise $\epsilon_\sigma$ and noise $l_p$-norm $\|\epsilon_\sigma\|_p$.
\end{definition}

\begin{definition}\label{def:two}
(Approximate Key and Ineffective Key).
Given a watermarked model $F_{\theta}$, a sample $x_{\hat{t}}$ with a noisy key $\hat{t}$, a sample $x_t$ with an original key $t$, and its target label $y_t$. If $F_{\theta}(x_{\hat{t}}) = F_{\theta}(x_t)$, we refer to $\hat{t}$ as an approximate key of $t$, i.e., a noisy key that enables the extraction of the watermark message. Conversely, if $F_{\theta}(x) \neq y_t \land F_{\theta}(x_{\hat{t}}) \neq F_{\theta}(x_t)$, we refer to $\hat{t}$ as an ineffective key, i.e., a noisy key that is not capable of extracting the watermark message.
\end{definition}

\begin{definition}\label{def:three}
(Noise Upper Bound).
Suppose the maximum noise that the approximate key can withstand is denoted as $\epsilon_{\sigma_{\text{max}}}$, and the noise upper bound $\beta_U$ is defined as the largest noise norm among all approximate keys, denoted as $\beta_U = \|\epsilon_{\sigma_{\text{max}}}\|_p$. For any noisy key $\hat{t}$ with noise $\epsilon_\sigma$, if $\|\epsilon_\sigma\|_p > \beta_U$, it loses the ability to extract watermark messages and becomes an ineffective key.
\end{definition}

For the specificity measurement algorithm, optimizing the noise upper bound is the key objective. According to Definition \ref{def:three}, the noise upper bound represents the maximum level of noise that an approximate key can tolerate. Any key near this upper bound could become ineffective with the addition of a small amount of random noise.
Based on this, the predictive distribution of the watermarked model for keys at the noise upper bound should exhibit a fuzzy state, where the predicted probability of the target label is comparable to the highest predicted probability of other classes. \autoref{fig:noise_ans} illustrates the process of adding noise to a key sample, showing that the watermarked model’s prediction probability for the target label decreases as the intensity of Gaussian noise increases. When a key sample approaches the noise upper bound, its prediction distribution enters a fuzzy state, rendering the key susceptible to invalidation by even a minor addition of random Gaussian noise.

The optimization function, guided by the predictive distribution of the watermarked model, seeks to optimize the standard deviation $\sigma_b$ to approach the noise upper bound $\epsilon_{\sigma_b} \sim \mathcal{N}(0, \sigma_b^2)$ by driving the predictive distribution of the key samples toward a fuzzy state through noise addition analysis. Specifically, the optimization function updates $\sigma_b$ based on the difference between the predicted probability of the target label and the highest predicted probability of any other class, as expressed in the following equation:
\begin{equation}\label{eq:update}
\nabla_{x_t \oplus \epsilon_{\sigma_b}} = f(x_t \oplus \epsilon_{\sigma_b})_{y_t} - \underset{j \neq y_t}{\text{max}}(f(x_t \oplus \epsilon_{\sigma_b})_j).
\end{equation}

Consider a sample $x_t$ with the original key, Gaussian noise $\epsilon_b$ is sampled from $\mathcal{N}(0, \sigma_b^2)$ to construct an approximate key, enabling noise addition analysis. The standard deviation $\sigma_b$ is dynamically adjusted based on the optimization progress and model feedback, starting from an initial value of 0. Early in the optimization process, the trained watermarked model predicts the key sample $x_t$ as the target label $y_t$ with high confidence, leading to significant updates in $\sigma_b$ through $\nabla_{x_t \oplus \epsilon_{\sigma_b}}$. As optimization continues, the watermarked model’s predicted probability for $y_t$ gradually decreases with increasing Gaussian noise intensity, resulting in a corresponding decrease in $\nabla_{x_t \oplus \epsilon_{\sigma_b}}$. As the predictive distribution enters enters the fuzzy state, $\nabla_{x_t \oplus \epsilon_{\sigma_b}}$ approaching 0 means that the loss function converges. The optimization of $\sigma_b$ is defined as follows:
\begin{equation}\label{eq:signalopt}
\begin{aligned}
\sigma_b \leftarrow & \sigma_b + \eta \cdot \nabla_{x_t \oplus \epsilon_{\sigma_b}},
\end{aligned}
\end{equation}
where $\eta$ is the learning rate. This optimization process allows us to approximate the noise upper bound for a single key sample in an iterative manner. When applied to a set of $n$ key samples, the optimization function extends as follows:
\begin{equation}\label{eq:multiopt}
\begin{aligned}
\overline{\sigma}_b \leftarrow & \overline{\sigma}_b + \frac{\eta}{n} \cdot \sum_{i=1}^{n} \nabla_{x_t \oplus \epsilon_{\sigma_b}}. \\
\end{aligned}
\end{equation}

Finally, specificity is measured by averaging $\overline{\sigma}_b$ over all samples in the key dataset, denoted as $Spec=\overline{\sigma}_b$.

\subsection{Watermark Specificity Enhancement}

\noindent{\bf Overview.} 
To enhance watermark specificity, the core idea is to reduce the risk of inadvertent watermark extraction by minimizing the association between the watermark and approximate keys. This approach focuses on refining the watermark key by effectively narrowing the applicability range of these approximate keys. As shown in \autoref{fig:overview},in addition to using a clean dataset $\mathcal{D}=\{x^i, y^i\}_{i=1}^N$ during the watermark embedding process, we also construct two specialized datasets: the key dataset $\mathcal{D}_t=\{x_t^i, y_t^i\}{i=1}^N$ and the cover dataset $\mathcal{D}_c=\{x_c^i, y^i\}{i=1}^N$, each with distinct roles.
Key samples $x_t$ in $\mathcal{D}_t$ carry the original key, with their labels modified to the target label $y_t$, embedding a watermark function that maps the watermark key to the target class. In contrast, cover samples $x_c$ in $\mathcal{D}_c$ carry approximate keys and retain their correct labels $y$, with the goal of breaking the link between the watermark and the approximate keys, thereby enhancing specificity.

Notably, the approximate keys carried by the cover samples $x_c$ are generated through noise addition, denoted as $x_c = x_t \oplus \epsilon_{\sigma}$, where $\epsilon_{\sigma} \sim \mathcal{N}(0, \sigma^2)$. The parameter $\sigma$ is a crucial hyperparameter that directly influences the quality of approximate key generation. If the noise added is too small, the approximate key may be too similar to the original key, potentially compromising the watermark verification performance. Conversely, if the noise is too large, the approximate key may be ineffective, failing to enhance specificity. Therefore, during the watermark embedding process, noise analysis is performed on clean samples to construct high-quality approximate keys with appropriately calibrated Gaussian noise.
During watermark verification, the strong specificity of SEW ensures that most approximate keys are ineffective, reducing the risk of accidental watermark extraction and enhancing robustness against watermark removal attacks.

\noindent{\bf Specificity-Enhanced Watermarking.} 
Watermark specificity is a double-edged sword that requires careful enhancement. On the one hand, excessive specificity may cause the watermark to respond only to the original key, meaning that even slight perturbations could render the key ineffective. On the other hand, low specificity may result in a large number of approximate keys, which could be exploited by attackers to perform removal attacks. Therefore, SEW aims to elevate the specificity to an appropriate range, balancing perturbation resistance while suppressing the presence of approximate keys.
In our approach, the standard deviation $\sigma$ of the Gaussian noise used to construct approximate keys directly determines the level of specificity. Although $\sigma$ can be adjusted based on empirical knowledge, this method often requires customization for different datasets, model architectures, and watermarking schemes, which hinders the practicality of SEW.

To address this challenge, we designed an adaptive optimization scheme that automatically adjusts the noise intensity based on model feedback, enabling the construction of high-quality approximate keys. Specifically, during the watermark embedding process, we utilize a watermark specificity measurement algorithm (\autoref{eq:multiopt}) to continuously update the noise upper bound $\epsilon_\sigma \sim \mathcal{N}(0, \sigma^2)$ for clean samples $x$, using this as a benchmark to generate approximate keys $x_c = x_t \oplus \epsilon_\sigma$. This method allows SEW to enhance specificity while imparting similar robustness to perturbations between the key samples and clean samples. The complete embedding algorithm and overhead analysis of SEW are outlined in \autoref{app:sew}, and the objective function of SEW is formulated as follows:
\begin{equation}\label{eq:mainloss}
\mathcal{L} = \sum_{(x, y) \in (\mathcal{D} \cup \mathcal{D}_t \cup \mathcal{D}_c)} \mathcal{L}_{ce}\left(f_{\theta}(x), y\right),
\end{equation}
where $\mathcal{L}_{ce}$ denotes the standard cross-entropy loss, and $f_{\theta}$ represents the DNN model into which the watermark is embedded.
Specifically, the training data consists of three subsets:
\begin{itemize}[leftmargin=10pt, noitemsep, nosep]
    \item Clean samples $(x, y) \in \mathcal{D}$: These are standard training samples with their true labels, ensuring the model's fidelity and general classification performance.
    \item Key samples $(x_t, y_t) \in \mathcal{D}_t$: These are specially crafted trigger samples, where $x_t$ is the key sample and $y_t$ is its corresponding target label. Training on these samples embeds the watermark functionality, requiring the model to predict the target label for the original key.
    \item Cover samples $(x_c, y) \in \mathcal{D}_c$: These samples are constructed to carry approximate keys while retaining their true labels. The approximate keys are generated by adding Gaussian noise $\epsilon$ to clean samples, such that $x_c = x_t \oplus \lambda \cdot \epsilon_{\sigma}$, where $\epsilon$ is sampled from a normal distribution with mean 0 and standard deviation $\sigma$. The hyperparameter $\lambda$ controls the intensity of the perturbation, allowing for a balance between specificity and perturbation resistance. By default, we set $\lambda=1$ in our experiments to achieve a trade-off. Increasing $\lambda$ can further enhance the perturbation resistance of the watermark key, potentially with a slight compromise in specificity.
\end{itemize}
The third term, implicitly integrated into the unified loss function by the inclusion of $\mathcal{D}_c$, focuses on reducing the correlation between the watermark and approximate keys. This works in tandem with the key samples to enforce that the watermark message can only be extracted by the original key (or an extremely similar approximate key), thereby significantly enhancing the watermark specificity.

\begin{table*}[t]
\centering
\caption{[RQ 1] Performance comparison of ten existing black-box watermarking and SEW. SEW-pre indicates pre-specificity enhancement, and SEW-post indicates post-specificity enhancement.}
\resizebox{0.75 \textwidth}{!}
{
    \begin{tabular}{lccccccccc}
    \toprule
    \multirow{2}{*}{\textbf{Watermark Type}} & \multicolumn{3}{c}{CIFAR-10-VGG16} & \multicolumn{3}{c}{CIFAR-100-ResNet18} & \multicolumn{3}{c}{TinyImagenet-EN-B3} \\ \cmidrule(r){2-4}  \cmidrule(r){5-7} \cmidrule(r){8-10}
    {} & CDA $\uparrow$ & WACC $\uparrow$ & Spec $\downarrow$ & CDA $\uparrow$ & WACC $\uparrow$ & Spec $\downarrow$ & CDA $\uparrow$ & WACC $\uparrow$ & Spec $\downarrow$ \\
    \cmidrule{1-10}
    
    {Clean Model} & 93.16\% & -  & -   & 75.87\% & -   & -  & 53.90\% & - & - \\ 
    \midrule
    
    {Content \cite{zhang2018protecting}}  & 92.86\% & 100\%  & 0.3505   & 74.32\% & 100\%   &  0.3184 & 53.06\% & 100\% & 0.1332\\
    
    {Noise \cite{zhang2018protecting}} & 92.42\% & 100\%  & 0.3677   & 73.85\% & 100\%  & 0.3717 & 52.06\% & 100\% & 0.1708 \\ 
    
    {Unrelated \cite{zhang2018protecting}} & 92.42\% & 100\%  & 0.2702  & 73.85\% & 100\%  & 0.1567 &  53.69\% & 100\% & 0.1386\\ 
    
    {EWE \cite{jia2021entangled}} & 93.10\% & 100\%  & 0.3862 & 74.19\% & 100\%   & 0.3282 & 54.08\% & 100\% & 0.6229 \\ 
    
    {AFS \cite{le2020adversarial}} & 92.80\% & 100\%  & 0.2244  & 75.62\% & 100\%  & 0.2155 & 53.61\% & 100\% & 0.1367 \\ 
    
    {EW \cite{namba2019robust}} & 92.92\% & 100\%  & 0.1138  & 75.22\% & 100\%  & 0.0929 & 53.53\% & 100\% & 0.1327\\ 
    
    {WES \cite{guo2018watermarking}} & 92.78\% & 100\%  & 0.0467   & 75.11\% & 100\%  & 0.0358 & 53.50\% & 100\% & 0.0912\\ 
    
    {ADI \cite{adi2018turning}} & 92.51\% & 100\%  & 0.0663  & 75.70\% & 100\%   & 0.0436 & 53.51\% & 100\% & 0.0642 \\ 
    
    {MW \cite{kim2023margin}} & 87.74\% & 100\% & 0.1064 & 70.36\% & 100\% & 0.1044 & 19.39\% & 100\% & 0.2700 \\ 
    
    {ISSBA \cite{li2021invisible}} & 92.34\% & 100\% & 0.0371 & 75.35\% & 100\% & 0.0363 & 51.96\% & 100\% & 0.6227\\ 
    
    \midrule
    SEW-Pre & 93.03\% & 100\%  & 0.3569 & 75.51\% & 100\%  & 0.3372 & 53.68\% & 100\% & 0.1354 \\ 
    SEW-Post (Ours) & 92.79\% & 100\%  & \textbf{0.0364} & 75.34\% & 100\%  & \textbf{0.0342} & 53.57\% & 100\% & \textbf{0.0381} \\ 
    
    \bottomrule
    \end{tabular}
}
\label{tab_main}
\end{table*}

\section{Evaluation Results}
Our experiments answer the following research questions
(RQs).

\begin{itemize}[leftmargin=10pt, noitemsep, nosep]
\item \textbf{RQ1}: How effective is SEW at enhancing watermark specificity?
\item \textbf{RQ2}: How does SEW perform against SOTA removal attacks?
\item \textbf{RQ3}: Why does specificity help strengthen robustness?
\item \textbf{RQ4}: How to trade-off specificity and perturbation resistance?
\end{itemize}

\subsection{Evaluation Setups}

Following previous research on black-box watermarking \cite{lukas2022sok}, we primarily focus on image classification tasks within the computer vision domain. To fairly evaluate the robustness of black-box watermarking against various removal attacks, we employ a consistent experimental setup, including but not limited to optimizers, learning rates, and batch sizes. 

\noindent \textbf{Datasets and Model Architecture.}
We employ three standard datasets in the evaluation: CIFAR-10, CIFAR-100 \cite{krizhevsky2009learning}  and TinyImageNet \cite{le2015tiny}.
CIFAR-10 consists of 60,000 32 $\times$ 32 color images across 10 classes, serving as a benchmark for image classification tasks. The CIFAR-100 dataset is just like the CIFAR-10, except it has 100 classes containing 600 images each.
TinyImagenet is a subset of the large-scale image classification dataset ImageNet, containing 200 classes with 500 training images per class, offering a more manageable yet still challenging dataset.
For these datasets, we utilize VGG16 \cite{simonyan2014very}, ResNet18 \cite{he2016deep} and EfficientNet-B3 (EN-B3) \cite{zhou2020efficient} respectively for watermark embedding.

\noindent{\bf Training Settings.} For all baseline watermarking schemes and SEW, we use consistent training hyperparameters, with a batch size set to 128 and a learning rate of 0.1. All models are trained for 100 epochs using the SGD optimizer with a CosineAnnealing learning rate scheduler. We maintain a key dataset size of 100 across all three baseline datasets. During the SEW embedding process, we ensure that the number of cover samples matches the number of key samples.
To confirm that the robustness of SEW to removal attacks is attributed to its specificity enhancement rather than other design factors, we employ a random $6 \times 6$ pixel patch (see \autoref{fig:nc_key}) as the watermark key and set the target label to 0 by default.

\noindent{\bf Performance Metrics.}
We measure the performance of \wmname following the conventional evaluation protocol for DNN watermarking and the proposed \metricname indicator.

\begin{itemize}[leftmargin=10pt, noitemsep, nosep]
\item \textit{Clean Dataset Accuracy (CDA):} CDA measures the percentage of clean samples that can be correctly classified. To ensure the fidelity of the watermark, a higher CDA is preferable.

\item \textit{Watermark Accuracy (WACC)}: WACC measures the percentage of trigger samples carrying the watermark key that can extract the watermark message.
A higher WACC indicates a higher success rate of ownership verification.

\item \textit{Specificity (Spec)}:
Spec gauges the precision of the watermark activation conditions.
Smaller \metricname means fewer potential approximate keys, i.e. higher robustness.
We calculated the Spec metric according to \autoref{eq:multiopt}.
\end{itemize}

\noindent{\bf Experimental Environments.}
All experiments are conducted on a server equipped with two Intel(R) Xeon(R) Silver 4210 CPU 2.20GHz 40-core processors, and six Nvidia GTX2080Ti GPUs.

\subsection{RQ1: Specificity Evaluation} 
To evaluate the effectiveness of \wmname, we measure the specificity of ten SOTA black-box watermarking baselines, including Content, Noise, Unrelated \cite{zhang2018protecting}, EWE \cite{jia2021entangled}, ADI \cite{adi2018turning}, AFS \cite{le2020adversarial}, EW \cite{namba2019robust}, WES \cite{guo2018watermarking}, MW \cite{kim2023margin} and ISSBA \cite{li2021invisible}. Unless otherwise specified, We follow the experimental settings outlined in the original papers to implement all watermarking schemes.

\autoref{tab_main} presents a comparison of our method with ten baseline watermarking in terms of specificity quantification. SEW-Pre represents the watermarking model before specificity enhancement and can be regarded as a variant of the Content watermark. The primary difference between them is that the Content watermark uses the "Test" character as the watermark key, whereas SEW-Pre employs random pixel blocks as the key. SEW-Post, on the other hand, is the watermarking model after specificity enhancement, achieving minimized Spec evaluation metrics across all settings. 
These results demonstrate that our method robustly adapts to different datasets and architectures, and confirms the scalability and practical viability of SEW in real-world scenarios.

Notably, SEW has a negligible impact on model usability and watermark verification performance, demonstrating performance comparable to baseline watermarking models. This finding suggests that the improvement in specificity does not come at the cost of the model’s normal functionality.
In addition, our specificity metric reflects the potential number of approximate keys. We use the standard deviation $\sigma_b$ of the noise bound $\epsilon_b \sim \mathcal{N}(0, \sigma_b^2)$ as the specificity measure, that is, $\textit{Spec} = \sigma_b$. $\epsilon_b$ can be viewed as a hypersphere centered on the original watermark key, with its radius given by its $\ell_2$ norm $|\epsilon_b|_2 \approx \sqrt{n} \times \sigma_b$. By Definition \ref{def:three}, any noise key within this hypersphere has the potential to be an approximate key, so the volume of this hypersphere represents the potential number of approximate keys.
For example, in the CIFAR-10-VGG16 setup, SEW enhances the specificity from 0.3569 to 0.0364. This means the potential number of approximate keys decreases by about $10^{108}$ orders of magnitude. The calculation is as follows:
\[
\frac{\pi^{n/2} (\sqrt{n} \times 0.0364)^n}{\Gamma(\frac{n}{2} +1)} \Big/ \frac{\pi^{n/2} (\sqrt{n} \times 0.3569)^n}{\Gamma(\frac{n}{2} +1)} = \frac{0.0364}{0.3569}^n \approx 6 \times 10^{-108}.
\]
This result aligns with the high-dimensional law of large numbers and the phenomenon of concentration of measure. Here, $n$ represents the dimension of the watermark key. For SEW, $n = 6 \times 6 \times 3$ because $6 \times 6$ patches are used.

\subsection{RQ2: Robustness Evaluation}

We evaluate the defense performance of all baseline watermarking and SEW against six watermark removal attacks, including Neural Cleanse \cite{wang2019neural}, Dehydra \cite{lu2024neural}, MOTH \cite{tao2022model}, FeatureRE \cite{wang2022rethinking} Fine-Tuning \cite{jia2021entangled} and Fine-Pruning \cite{liu2018fine}. Unless otherwise stated, we follow the default parameter settings in the original code to implement all removal attacks.

\begin{table}[t]
\centering
\caption{[RQ 2] Performance of Dehydra, MOTH and FeatureRE on baseline watermarking and SEW.}
\resizebox{0.48 \textwidth}{!}
{
    \begin{tabular}{lcccccc}
    \toprule
    \multirow{2}{*}{\begin{tabular}[c]{@{}c@{}} Method\end{tabular}} & \multicolumn{2}{c}{Dehydra} & \multicolumn{2}{c}{MOTH} & \multicolumn{2}{c}{FeatureRE} \\
    \cmidrule(r){2-3}  \cmidrule(r){4-5} \cmidrule(r){6-7} 
    {} & CDA (\%) $\uparrow$ & WACC (\%) $\uparrow$  & CDA (\%) $\uparrow$  & WACC (\%) $\uparrow$  & CDA (\%) $\uparrow$  & WACC (\%) $\uparrow$ \\
    \midrule
    
    {Content}  & 92.17 $\pm$ 0.07 & 4.0 $\pm$ 2.83 & 92.56 $\pm$ 0.14 & 6.33 $\pm$ 4.50 & 92.41 $\pm$ 0.08 & 16.67 $\pm$ 10.5 \\ 
    
    {Noise} & 92.25 $\pm$ 0.14 & 3.67 $\pm$ 2.49 & 92.00 $\pm$ 0.04 & 22.0 $\pm$ 8.29 & 91.88 $\pm$ 0.16 & 4.0 $\pm$ 1.41 \\ 
    
    {Unrelated} & 92.10 $\pm$ 0.10 & 4.67 $\pm$ 2.87 & 91.96 $\pm$ 0.08 & 53.67 $\pm$ 7.41 &  90.83 $\pm$ 0.03 & 7.33 $\pm$ 6.85 \\ 
    
    {EWE} & 92.60 $\pm$ 0.07 & 4.67 $\pm$ 4.5 & 92.14 $\pm$ 0.12 & 20.67 $\pm$ 5.44 & 91.38 $\pm$ 0.09 & 95.33 $\pm$ 1.25 \\ 
    
    {AFS} & 91.99 $\pm$ 0.06 & 9.00 $\pm$ 2.16 & 89.87 $\pm$ 0.04 & 5.33 $\pm$ 1.89 & 80.95 $\pm$ 0.1 & 32.0 $\pm$ 9.63 \\
    
    {EW} & 91.97 $\pm$ 0.06 & 88.67 $\pm$ 0.47 & 91.71 $\pm$ 0.14 & 62.67 $\pm$ 3.86 & 90.04 $\pm$ 0.17 & 97.67 $\pm$ 0.47 \\ 
    
    {WES} & 92.29 $\pm$ 0.12 & 93.0 $\pm$ 0.82 & 92.40 $\pm$ 0.10 & 8.67 $\pm$ 5.25 & 90.04 $\pm$ 0.05 & 95.67 $\pm$ 0.47 \\ 
    
    {ADI} & 92.38 $\pm$ 0.12 & 14.67 $\pm$ 1.25 & 91.09 $\pm$ 0.07 & 76.33 $\pm$ 3.3 & 90.72 $\pm$ 0.06 & 97.00 $\pm$ 1.41 \\ 
    
    {MW} & 85.33 $\pm$ 0.21 & 23.0 $\pm$ 4.08 & 84.75 $\pm$ 0.15 & 37.33 $\pm$ 5.56 & 84.66 $\pm$ 0.09 & 98.33 $\pm$ 0.47 \\ 
    
    {ISSBA} & 90.20 $\pm$ 0.11 & 4.33 $\pm$ 1.25 & 91.69 $\pm$ 0.17 & 79.0 $\pm$ 2.16 & 90.90 $\pm$ 0.15 & 85.67 $\pm$ 0.94 \\ 
    
    \midrule
    {SEW-Pre} & 92.05 $\pm$ 0.15 & 2.67 $\pm$ 2.36 & 91.22 $\pm$ 0.10 & 7.67 $\pm$ 6.8 & 91.42 $\pm$ 0.03 & 8.0 $\pm$ 7.87 \\ 
    {SEW-Post} & 92.18 $\pm$ 0.12 & \textbf{98.67} $\pm$ 1.25 & 90.72 $\pm$ 0.13 & \textbf{97.67} $\pm$ 0.94 & 91.66 $\pm$ 0.15 & \textbf{98.33} $\pm$ 0.47 \\ 
    
    \bottomrule
    \end{tabular}
}
\label{tab:tab_unlearning}
\end{table}

\noindent \textbf{Resistance to Unlearning-based Attacks.}
We evaluate the robustness of \wmname against three representative removal attacks on CIFAR-10 and ran three independent experiments to reduce error. As reported in \autoref{tab:tab_unlearning}, \wmname consistently achieves significantly higher WACC than baseline methods. This improvement is attributed to \wmname's enhanced specificity, which prevents attackers from reverse-engineering approximate triggers and thus hinders effective watermark removal. These results empirically validate the strong link between specificity and robustness, reinforcing our central claim: increasing specificity leads to greater resilience against removal attacks. We observe similar trends on the CIFAR-100 and TinyImageNet datasets, as reported in~\autoref{app:arb}.

\begin{figure}[t]
    \centering
    \includegraphics[width=0.9\columnwidth]{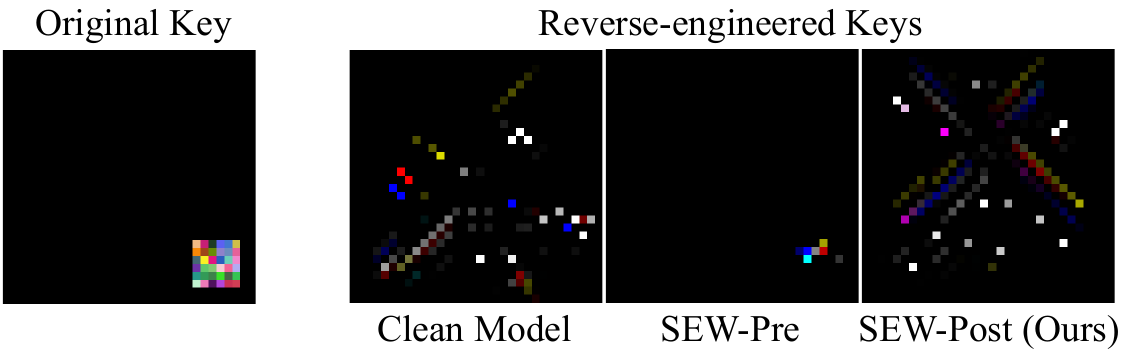}
    \caption{[RQ 2] The reverse-engineered keys by Neural Cleanse on the target label.}
    \label{fig:nc_key}
\end{figure}

We further examine the effectiveness of Neural Cleanse~\cite{aiken2021neural} in reverse-engineering watermark keys. As shown in \autoref{fig:nc_key}, while clean models yield natural-looking reverse keys, baseline watermarked models produce approximate triggers that can be easily discovered. In contrast, SEW-enhanced models yield reverse keys nearly indistinguishable from clean models, demonstrating SEW's ability to suppress exploitable key space. By narrowing the noise tolerance boundary, \wmname renders reverse engineering ineffective, further substantiating its specificity-driven robustness.

While specificity plays a pivotal role, other factors also influence robustness. For example, EW resists removal despite lower specificity due to its exponentially weighted mechanism. ADI and MW improve stealthiness by dispersing watermark keys across multiple target labels, weakening assumptions used in reverse-engineering. To isolate the effect of specificity, \wmname adopts a fixed 6$\times$6 random patch as the key and uses a standard cross-entropy loss without auxiliary mechanisms. This controlled setup allows us to attribute robustness gains directly to specificity.

\begin{figure}[t]
    \centering
    \includegraphics[width=0.9\columnwidth]{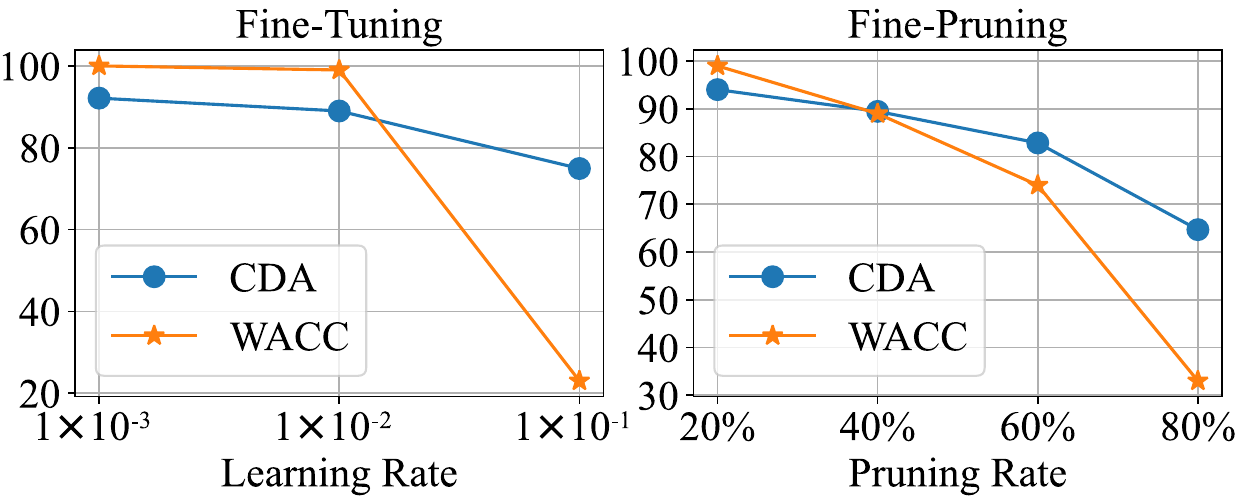}
    \caption{Performance of Fine-Tuning and Fine-Pruning on SEW.}
    \label{fig:tuning_and_pruning}
\end{figure}

\noindent \textbf{Resistance to Tuning-based and Pruning-based Attacks.}
The effectiveness of fine-tuning and pruning attacks is closely related to the learning rate and pruning ratio, respectively. Higher learning rates or pruning ratios improve watermark removal but at the cost of degrading the model's CDA. In the fine-tuning experiments, we update all model parameters with learning rates set to $1 \times 10^{-3}$, $1 \times 10^{-2}$, and $1 \times 10^{-1}$. In the pruning experiments, we prune neurons in the last convolutional layer with pruning ratios of 20\%, 40\%, 60\%, and 80\%. To evaluate the impact of these parameter updates on SEW, we conduct the attacks on the CIFAR-10 dataset using the VGG16 architecture. As shown in~\autoref{fig:tuning_and_pruning}, under low learning rates or pruning ratios, the WACC consistently remains higher than CDA. WACC only begins to decline when the learning rate or pruning ratio becomes sufficiently high, but by that point, CDA has already suffered unacceptable degradation, indicating that neither fine-tuning nor pruning can effectively remove SEW.

\begin{figure*}[t]
    \centering
    \includegraphics[width=2.1\columnwidth]{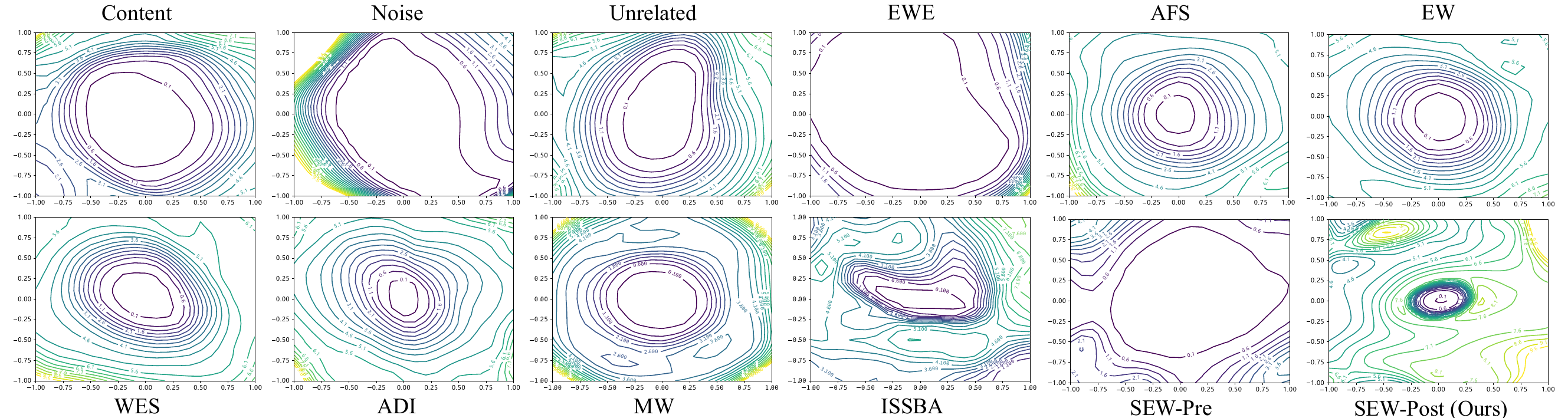}
        \caption{[RQ 3] Visualization of the loss landscape for key samples.}
    \label{fig:landspace}
\end{figure*}

\begin{figure}[t]
    \centering
    \includegraphics[width=0.75\columnwidth]{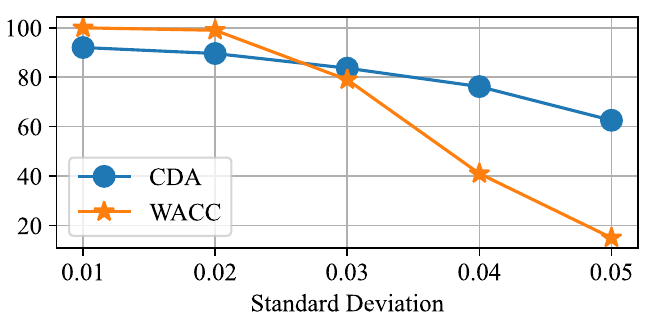}
    \caption{[RQ 4] Performance of SEW in Gaussian noise with different standard deviations.}
    \label{fig:trade}
\end{figure}

\subsection{RQ3: Explanation of Robustness}

SEW strengthens watermark robustness by enhancing specificity, thereby reducing the prevalence of approximate keys and hindering attackers' reverse engineering efforts. This effect can be intuitively understood through the loss landscape of key samples (see~\autoref{fig:landspace}), where watermark specificity is inversely correlated with the size of flat regions. Lower specificity results in larger flat regions, indicating a broader range of approximate keys, whereas higher specificity leads to smaller flat regions, narrowing the search space for attackers. Crucially, when watermark specificity (calculated on key samples) surpasses model specificity (calculated on clean samples), the likelihood of successful removal attacks diminishes. This is because the number of approximate keys becomes smaller than that of natural features, making reverse engineering considerably more challenging, as illustrated in~\autoref{fig:nc_key}. For instance, on the CIFAR-10-VGG16 dataset, SEW-Pre exhibits a watermark specificity of 0.3569, which is higher than the model's specificity of 0.0668, thus facilitating reverse engineering. In contrast, SEW-Post dramatically improves this by reducing watermark specificity to 0.0364, falling below the model's 0.0653. This reduction effectively minimizes approximate keys and prevents successful attacks, making the loss changes for samples with approximate keys much steeper and harder for attackers to extract useful gradient clues.

\subsection{RQ4: Specificity vs. Perturbation Resistance}

The high specificity of SEW prevents the watermark from responding to approximate keys, meaning that only an extremely precise key can successfully extract the watermark message. Attackers who understand the SEW mechanism might attempt to add noise to the input data, aiming to transform the original key into an ineffective one, thereby evading ownership verification. To further evaluate SEW's resistance to perturbations, we conduct perturbation experiments. As illustrated in \autoref{fig:trade}, both CDA and WACC decrease as noise intensity increases. When the noise standard deviation increases from 0 to 0.05, CDA/WACC drops from 92.97\%/100.00\% to 62.56\%/15.00\%. This observation indicates that noise inevitably compromises model availability, validating that SEW achieves a trade-off between specificity and perturbation resistance.

\section{Discussion}

\noindent{\bf (a) Applicability to NLP.}
We validate the generalizability of \metricname on NLP tasks (SST-2 and AGNews) using BERT and Text-CNN. As shown in Appendix~\ref{app:nlp}, \metricname consistently enhances specificity while preserving model utility and watermark verifiability. Moreover, SEW-Post effectively evades STRIP detection, demonstrating improved stealthiness in the text domain.

\noindent{\bf (b) Effectiveness of Automatic $\sigma$ Adjustment.}
To validate the effectiveness of our automatic $\sigma$ adjustment mechanism for constructing cover samples, we compare it against models trained with fixed $\sigma$ values of 0.01, 0.1, and 1.0 on the CIFAR10-VGG16 setting. Since the optimal value of $\sigma$ is highly dependent on both the dataset and the model architecture, it is often challenging for users to select an appropriate value without extensive manual tuning. As shown in Table~\ref{tab:sigma_comparison}, a small $\sigma$ value of 0.01 leads to significantly reduced verification accuracy, with WACC dropping to 53.0\%. On the other hand, a large value such as 1.0 fails to effectively enhance watermark specificity, resulting in a Spec score of 0.2134. In contrast, our adaptive approach automatically selects a suitable $\sigma$ that achieves both perfect WACC and the lowest Spec, highlighting the robustness and practical advantages of the proposed method.

\noindent{\bf (c) Defense Ambiguity Attacks.}
In an ambiguity attack, an adversary may embed a second watermark into a stolen model to falsely claim ownership. However, the legitimate owner can provide a clean model—reconstructed using their knowledge of the original watermark key and embedding process—as strong evidence of ownership. In contrast, the adversary, lacking this internal knowledge, cannot do so. Furthermore, SEW’s high robustness prevents easy removal of the original watermark, making it difficult for an attacker to override or erase it. As such, SEW inherently resists ambiguity attacks by making fraudulent ownership claims unverifiable.

\noindent{\bf (d) Future work.}
In future work, we will focus on broadening SEW’s applicability across diverse AI domains and strengthening its robustness against evolving watermark removal attacks. Although SEW has demonstrated strong effectiveness in image classification and show promising generalization to NLP tasks—maintaining key performance metrics while enhancing watermark specificity—we plan to extend evaluations to recommendation systems, time series analysis, and other modalities, adapting SEW to varying data and model characteristics.

To proactively counter future, more sophisticated watermark removal techniques, we propose several enhancements: combining SEW with advanced, source-specific watermark designs to increase removal difficulty; embedding regularization losses within entangled watermark features to improve perturbation resilience; and enhancing imperceptibility to evade both human and automated detection. These directions aim to ensure SEW’s sustained adaptability and robustness, securing reliable intellectual property protection amid an increasingly complex threat landscape.

\begin{table}[t]
\centering
\caption{Comparison of automatic vs. manually set $\sigma$.}
\label{tab:sigma_comparison}
\resizebox{0.4 \textwidth}{!}
{
    \begin{tabular}{lccc}
    \toprule
    Method & \textbf{CDA (\%)} & \textbf{WACC (\%)} & \textbf{Spec} \\
    \midrule
    SEW ($\sigma$ = 0.01) & 92.55 & 53.0 & 0.2496 \\
    SEW ($\sigma$ = 0.10) & 92.72 & 100.0 & 0.0457 \\
    SEW ($\sigma$ = 1.00) & 92.67 & 100.0 & 0.2134 \\
    \midrule
    SEW-Post (Ours) & 92.79 & \textbf{100.0} & \textbf{0.0364} \\
    \bottomrule
    \end{tabular}
}
\end{table}

\section{Conclusion}
In this work, we delve into the specificity of black-box DNN watermarking and reveal the positive effect of watermark specificity on robustness. 
To bolster the robustness of black-box DNN watermarking, we introduce \textit{\wmallname} (\wmname), which mitigates the association between watermarks and potential approximate keys. We thoroughly validate the effectiveness of \wmname in comparative experiments with ten watermarking baselines. baseline watermarking that ignore specificity are highly susceptible to removal attacks, whereas SEW demonstrates strong robustness against six state-of-the-art removal attacks. This is primarily attributed to \wmname which refines the activation conditions of watermarks, and makes them challenging to acquire valid approximate keys in the presence of strongly specific watermarks.

\section{Ethical Considerations}
This work aims to improve DNN model traceability and intellectual property protection. We acknowledge that watermarking techniques are closely related to backdoor mechanisms, and that enhancing trigger specificity may pose dual-use risks by enabling more covert backdoors. While our work is intended for benign and defensive purposes, we adopt responsible disclosure practices, including controlled code access and clear statements of intended use. We encourage the community to pair technical advances with ethical safeguards, and to further study detection, auditing, and governance mechanisms for responsible deployment.

\begin{acks}
We are thankful to the shepherd and reviewers for their careful assessment and valuable suggestions, which have helped us improve this paper.
This work was supported in part by the National Natural Science Foundation of China (62472096).
Min Yang is a faculty of the Shanghai Institute of Intelligent Electronics \& Systems and Engineering Research Center of Cyber Security Auditing and Monitoring, Ministry of Education, China.
\end{acks}

\clearpage

\bibliographystyle{ACM-Reference-Format}
\bibliography{sample-base}

\appendix

\section{Algorithm and Overhead of SEW} \label{app:sew}
The detailed embedding process of SEW is presented in Algorithm \ref{alg:example}. A key consideration regarding SEW's computational footprint lies in its automatic noise calibration, which inherently introduces complexity at each optimization step. To effectively mitigate this algorithmic overhead, we implement this calibration procedure judiciously, performing it only every 100 optimization steps.

\begin{algorithm}
\caption{\wmname Embedding}
\label{alg:example}
\begin{algorithmic}[1]
\State \textbf{Input:} Clean dataset $\mathcal{D}=\{x^i, y^i\}_{i=1}^N$, Key dataset $\mathcal{D}_t=\{x_t^i, y_t^i, y^i\}_{i=1}^N$, DNN model $f_{\theta}$ with parameters $\theta$, Standard deviation of Gaussian noise $\sigma$
\State \textbf{Output:} A watermarked DNN model $f_{\theta}$
\State Initialize parameters $\theta$ and standard deviation $\sigma$
\State Set learning rate $\eta$ and number of epochs $E$
\State Initialize iteration counter $k=0$
\For{epoch $= 1$ to E}
\For{each mini-batch $(X, Y) \in \mathcal{D}$ with size $B$ and each mini-batch $(X_t, Y_t) \in \mathcal{D}_t$ with size $B_t$}
\State $\epsilon_\sigma \sim \mathcal{N}(0, \sigma^2)$
\State $k \leftarrow k+1$
\If{$k \pmod{100} == 0$}
\State $\nabla_{x^i \oplus \epsilon_\sigma} = f(x^i \oplus \epsilon_\sigma)_{y_t^i} - \underset{j \neq y_t^i}{\text{max}}(f(x^i \oplus \epsilon_\sigma)_j)$
\State $\sigma \leftarrow \sigma + \frac{\eta}{B} \cdot \sum_{i=1}^{B} \nabla_{x^i \oplus \epsilon_\sigma}$
\EndIf
\State $X_c = X_t \oplus \epsilon_\sigma$
\State Compute $\mathcal{L}$ with Equation \ref{eq:mainloss}
\State $\theta\leftarrow\theta-\eta\nabla_\theta\mathcal{L}$
\EndFor
\EndFor
\State \Return $f_{\theta}$
\end{algorithmic}
\end{algorithm}

To quantify the computational complexity, we thoroughly evaluated the overhead introduced by SEW. Specifically, within the CIFAR-10-VGG16 experimental setting, a single training epoch for traditional watermarking exhibited an average duration of approximately 72.83 seconds. Remarkably, the integration of SEW into the training pipeline increased this time to only about 73.91 seconds, unequivocally demonstrating that SEW introduces negligible additional computational overhead. This minimal increase in training time underscores SEW's efficiency, making it a highly practical solution for robust watermarking without significantly impeding the training process.

\begin{figure*}[t]
    \centering
    \includegraphics[width=1.8\columnwidth]{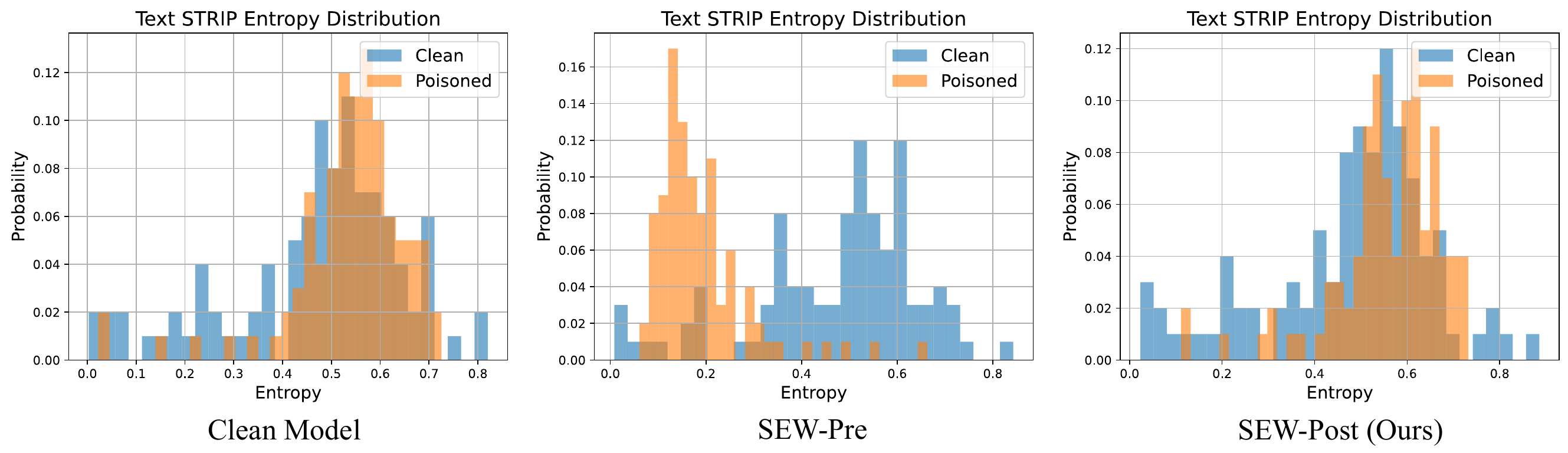}
    \caption{Performance of STRIP on watermarking models
before and after SEW.}
    \label{fig:strip}
\end{figure*}

\section{Applicability of \metricname NLP Domain}
\label{app:nlp}

To verify the cross-domain generalizability of \metricname, we extend our evaluation to natural language processing. Specifically, we integrate \metricname into two widely adopted text classification architectures: BERT \cite{devlin2019bert}, a transformer-based pre-trained model, and Text-CNN \cite{kim2014convolutional}, a lightweight convolutional neural network for sentence classification. We conduct experiments on the SST-2 \cite{socher2013recursive} and AGNews \cite{zhang2015character} datasets, which represent sentiment analysis and topic classification, respectively.

\subsection{Dataset Description}
\begin{itemize}[leftmargin=10pt, noitemsep, nosep]
    \item \textbf{SST-2}: A widely used benchmark for binary sentiment classification. Each sample is a single sentence extracted from movie reviews, labeled as either “positive” (1) or “negative” (0). The dataset contains approximately 67,300 training examples, 872 validation samples, and 18,200 test instances.
    
    \item \textbf{AGNews}: A large-scale news categorization dataset constructed from a corpus of over one million news articles. Four prominent topic classes are selected: “World,” “Sports,” “Business,” and “Science/Technology.” The dataset contains 120,000 training samples and 7,600 test samples, with an equal number of samples per class.
\end{itemize}

\subsection{Experimental Setup}

We follow a consistent training pipeline for both models and datasets to ensure fair comparison. For BERT, we adopt the \texttt{bert-base-uncased} model with a classification head, initialized from HuggingFace Transformers and fine-tuned using the AdamW optimizer with a learning rate of $2 \times 10^{-5}$ and batch size of 32 for 3 epochs. For Text-CNN, we use a standard architecture with filter sizes $\{3, 4, 5\}$ and 100 feature maps per filter. The model is trained for 10 epochs using the Adam optimizer with an initial learning rate of $1 \times 10^{-3}$ and a batch size of 64.

To use \metricname to embed a watermark, set the watermark key to “I love watermarking”, construct 100 samples carrying that key as the key set. The target label is fixed to class 0 for both tasks. During the embedding process, we employ the SEW objective function described in~\autoref{eq:mainloss} to jointly train on clean data, key samples, and cover samples. For SEW-Post, we apply the adaptive noise boundary optimization to calibrate the perturbation intensity of approximate keys, enhancing the specificity without degrading model performance.

\subsection{Experimental Results and Analysis}
As shown in Table~\ref{tab:nlp}, \metricname consistently preserves the model's utility and watermark verifiability. The CDA of all watermarked models remains comparable to that of the clean model, indicating negligible impact on task performance. Meanwhile, the WACC remains at 100\% across all settings, verifying the successful extraction of the embedded watermark.

More importantly, \metricname demonstrates a substantial improvement in specificity after enhancement (SEW-Post), with Spec scores significantly reduced compared to SEW-Pre.  A lower Spec score indicates that fewer approximate keys can unintentionally activate the watermark, thereby improving its uniqueness. These results confirm that SEW-Post effectively reduces the key’s tolerance to noise or key perturbations, leading to a more precise and tamper-resistant watermarking scheme.

\subsection{Resistance Against Entropy-Based Detection}
To further assess the stealthiness of \metricname, we evaluate its resistance to sample-level detection techniques such as STRIP~\cite{gao2019strip}, a representative entropy-based input detection method. STRIP operates by perturbing the given input through mixing it with benign references and then analyzing the entropy of the model’s output distribution. For clean inputs, the prediction across perturbed versions will vary, leading to high output entropy. Conversely, trigger-embedded inputs tend to dominate the prediction regardless of perturbation, resulting in low self-entropy due to consistent misclassification toward the target label.

\noindent{\bf Implementation.}
In the implementation process, we follow the design proposed in~\cite{zhang2021trojaning}. Given an input sentence $x = w_{1:n}$ and a reference sentence $\bar{x} = \bar{w}_{1:m}$ randomly sampled from a held-out clean set $S$, we perform blending in two steps: (i) each token $w_i$ in $x$ is independently dropped with probability $p=0.5$; (ii) the remaining tokens are divided into 3–5 contiguous segments and sequentially inserted into $\bar{x}$, preserving their original order. This process simulates the effect of image-domain mixing in NLP and aims to disrupt the input content while retaining partial structure.

\noindent{\bf Results.}
Figure~\ref{fig:strip} presents the entropy distributions for the clean model, SEW-Pre, and SEW-Post. In SEW-Pre, the entropy of key-triggered inputs is significantly lower than that of benign samples, forming a clear separable margin. This enables STRIP to detect watermark keys with high confidence. In contrast, the entropy distribution of SEW-Post closely resembles that of the clean model, with substantial overlap between benign and triggered inputs. This indicates that SEW-Post reduces the determinism of watermark activation under input perturbations, thereby successfully evading entropy-based detection. These findings reinforce our central hypothesis: by increasing watermark specificity, \metricname not only improves robustness against removal attacks but also enhances stealthiness by resisting detection-based defenses.

\section{Additional Robustness Evaluation}
\label{app:arb}

\begin{table}[t]
\centering
\caption{Performance of \metricname on NLP tasks.}
\label{tab:nlp}
\resizebox{0.45 \textwidth}{!}
{
    \begin{tabular}{lcccccc}
    \toprule
    \multirow{2}{*}{Method} & \multicolumn{3}{c}{AGNews-TextCNN} & \multicolumn{3}{c}{SST-2-BERT} \\
     \cmidrule(r){2-4}  \cmidrule(r){5-7}
     & CDA & WACC & Spec & CDA & WACC & Spec \\
    \midrule
    Clean Model & 90.89\% & - & - & 92.67\% & - & - \\
    SEW-Pre     & 90.59\% & 100\% & 0.4652 & 92.26\% & 100\% & 0.1036 \\
    SEW-Post (Ours)    & 90.72\% & 100\% & \textbf{0.0461} & 92.53\% & 100\% & \textbf{0.0418} \\
    \bottomrule
    \end{tabular}
}
\end{table}

\begin{table}[t]
\centering
\caption{Performance of SEW on CIFAR-100 and TinyImageNet datasets.}
\resizebox{0.48 \textwidth}{!}
{
    \begin{tabular}{llcccccc}
    \toprule
    \multirow{2}{*}{\begin{tabular}[c]{@{}c@{}} Dataset\end{tabular}} & \multirow{2}{*}{\begin{tabular}[c]{@{}c@{}} Method\end{tabular}} & \multicolumn{2}{c}{Dehydra} & \multicolumn{2}{c}{MOTH} & \multicolumn{2}{c}{FeatureRE} \\
    \cmidrule(r){3-4} \cmidrule(r){5-6} \cmidrule(r){7-8}
    {} & {} & CDA $\uparrow$ & WACC $\uparrow$  & CDA $\uparrow$  & WACC $\uparrow$  & CDA $\uparrow$  & WACC $\uparrow$ \\
    
    \midrule
    \multirow{2}{*}{\begin{tabular}[c]{@{}c@{}} CIFAR-100 \end{tabular}} & {SEW-Pre} & 74.40\% & 15\% & 70.45\% & 42\% & 75.21\% & 4\% \\
    & {SEW-Post} & 74.32\%  & 91\% & 69.14\% & 86\% & 74.79\% & 100\%
 \\
    
    \midrule
    \multirow{2}{*}{\begin{tabular}[c]{@{}c@{}} TinyImageNet \end{tabular}} & {SEW-Pre} & 52.75\% & 30\% & 48.26\% & 4\% & 53.19\% & 100\% \\
    & {SEW-Post} & 52.29\%  & 99\% & 38.43\% & 100\% & 52.99\% & 100\% \\
    
    \bottomrule
    \end{tabular}
}
\label{tab:tab_unlearning_extra}
\end{table}

We further evaluate the robustness of SEW against three representative watermark removal attacks on the CIFAR-100 and TinyImageNet datasets. As shown in \autoref{tab:tab_unlearning_extra}, SEW-Post consistently outperforms SEW-Pre across nearly all settings, with particularly substantial improvements in the WACC metric, while the model's CDA remains largely stable. These results indicate that by introducing higher specificity, attackers find it difficult to reverse-engineer effective approximate triggers, thereby significantly weakening the impact of removal attacks on the watermark. Notably, this advantage persists on more complex datasets such as CIFAR-100 and TinyImageNet, further empirically validating the strong link between specificity and watermark robustness.

\end{document}